\documentclass[%
twocolumn,
amsmath,amssymb,
aps,
superscriptaddress,
]{revtex4-2}

\usepackage[svgnames,psnames]{xcolor}
\usepackage{url}
\usepackage{graphicx}
\graphicspath{{.}{figs/}{../figs/}}
\usepackage{afterpage}
\usepackage{siunitx}
\DeclareSIUnit\angstrom{\text {Å}}
\usepackage{dcolumn}
\usepackage{bm}
\usepackage{tikz}
\usetikzlibrary{positioning,intersections,angles,quotes}
\usepackage[caption=false]{subfig}
\usepackage{booktabs} 
\usepackage{comment}
\usepackage{physics}
\usepackage{mathtools}

\usepackage[colorlinks,citecolor=DarkGreen,linkcolor=FireBrick,linktocpage,unicode,hypertexnames=false]{hyperref}
\usepackage[capitalize]{cleveref}
\crefname{enumi}{}{}
\usepackage{autonum}

\begin{document}

\title{Quantifying fermionic nonlinearity of quantum circuits}

\author{Shigeo Hakkaku}
\email{shigeo.hakkaku@ntt.com}
\affiliation{%
  Graduate School of Engineering Science, Osaka University, 1-3 Machikaneyama, Toyonaka, Osaka 560-8531, Japan
}%
\affiliation{%
  NTT Computer and Data Science Laboratories, NTT Corporation, 3-9-11 Midoricho, Musashino 180-8585, Japan
}%
\author{Yuichiro Tashima}
\email{u752467b@gmail.com}
\affiliation{%
  Graduate School of Engineering Science, Osaka University, 1-3 Machikaneyama, Toyonaka, Osaka 560-8531, Japan
}%
\author{Kosuke Mitarai}
\email{mitarai.kosuke.es@osaka-u.ac.jp}
\affiliation{%
  Graduate School of Engineering Science, Osaka University, 1-3 Machikaneyama, Toyonaka, Osaka 560-8531, Japan
}%
\affiliation{%
  Center for Quantum Information and Quantum Biology,
  Osaka University, 1-2 Machikaneyama, Toyonaka 560-0043, Japan
}%
\affiliation{%
  JST,
  PRESTO,
  4-1-8 Honcho, Kawaguchi, Saitama 332-0012, Japan
}%
\author{Wataru Mizukami}
\email{mizukami.wataru.qiqb@osaka-u.ac.jp}
\affiliation{%
  Graduate School of Engineering Science, Osaka University, 1-3 Machikaneyama, Toyonaka, Osaka 560-8531, Japan
}%
\affiliation{%
  Center for Quantum Information and Quantum Biology,
  Osaka University, 1-2 Machikaneyama, Toyonaka 560-0043, Japan
}%
\affiliation{%
  JST,
  PRESTO,
  4-1-8 Honcho, Kawaguchi, Saitama 332-0012, Japan
}%
\author{Keisuke Fujii}%
\email{fujii@qc.ee.es.osaka-u.ac.jp}
\affiliation{%
   Graduate School of Engineering Science, Osaka University, 1-3 Machikaneyama, Toyonaka, Osaka 560-8531, Japan
}%
\affiliation{%
  Center for Quantum Information and Quantum Biology,
  Osaka University, 1-2 Machikaneyama, Toyonaka 560-0043, Japan
}%
\affiliation{%
  RIKEN Center for Quantum Computing (RQC),
  Hirosawa 2-1, Wako, Saitama 351-0198, Japan
}%
\affiliation{%
  Fujitsu Quantum Computing Joint Research Division at QIQB,
  Osaka University, 1-2 Machikaneyama, Toyonaka 560-0043, Japan
}%
\date{\today}
\begin{abstract}
Variational quantum algorithms (VQAs) have been proposed as 
one of the most promising approaches to demonstrate quantum advantage on noisy intermediate-scale quantum (NISQ) devices.
However, it has been unclear whether VQAs can maintain quantum advantage under the intrinsic noise of the NISQ devices, which deteriorates the quantumness.
Here we propose a measure, called \textit{fermionic nonlinearity}, to quantify the classical simulatability of quantum circuits designed for simulating fermionic Hamiltonians.
Specifically, we construct a Monte Carlo type classical algorithm based on the classical simulatability of fermionic linear optics,
whose sampling overhead is characterized by the fermionic nonlinearity.
As a demonstration of these techniques, we calculate the upper bound of the fermionic nonlinearity of a rotation gate generated by four fermionic modes under the dephasing noise.
Moreover, we estimate the sampling costs of the unitary coupled cluster singles and doubles quantum circuits for hydrogen chains subject to the dephasing noise.
We find that, depending on the error probability and atomic spacing, there are regions where the fermionic nonlinearity becomes very small or unity, and hence the circuits are classically simulatable.
We believe that our method and results help to design quantum circuits for fermionic systems with potential quantum advantages.
\end{abstract}

\maketitle

\section{Introduction}
\NewDocumentCommand\ansatz{}{\textit{Ansatz}}
\NewDocumentCommand\ansatze{}{\textit{Ans\"{a}tze}}
Quantum computers have attracted much attention because of their capability to solve classically intractable problems \cite{lloydUniversalQuantumSimulators1996,shorPolynomialTimeAlgorithmsPrime1997,harrowQuantumAlgorithmLinear2009}.
Among them, the first industrial application is expected to be a quantum chemistry calculation, which uses quantum computers for simulating fermionic many-body systems.
It has been predicted that a fault-tolerant quantum computer with about a million physical qubits can simulate both Fermi-Hubbard and molecular electronic structure Hamiltonians beyond classical approaches \cite{babbushEncodingElectronicSpectra2018}.
The application has also been anticipated for NISQ devices through the variational quantum eigensolver (VQE) \cite{peruzzoVariationalEigenvalueSolver2014,kandalaHardwareefficientVariationalQuantum2017,matsuzawajastrowtypeDecompositionQuantum2020,wiersemaExploringEntanglementOptimization2020,liEfficientVariationalQuantum2017,heyaSubspaceVariationalQuantum2019,hempelQuantumChemistryCalculations2018,namGroundstateEnergyEstimation2020}.

It is essential to predict at what scale quantum computers can have advantages over classical ones for those applications.
One way is to estimate the computational cost required for fault-tolerant quantum computers to simulate fermionic systems that are well beyond the reach of classical supercomputers \cite{babbushEncodingElectronicSpectra2018,kivlichanImprovedFaultTolerantQuantum2020,mottaLowRankRepresentations2021}.
However, this approach highly depends both on the objective system and algorithms employed in classical or quantum computers.
As another approach, we can ask a question the other way around; given a quantum circuit that simulates a fermionic system, what is the cost of classical computation to simulate that circuit?
If there exists a quantum advantage, at least such a quantum circuit has to be hard for a classical computer to be simulated.

One way to evaluate the classical simulatability of quantum circuits is to quantify the simulation cost of a specific quasiprobability-based simulator \cite{pashayanEstimatingOutcomeProbabilities2015,howardApplicationResourceTheory2017,seddonQuantifyingMagicMultiqubit2019,rallSimulationQubitQuantum2019,hakkakuComparativeStudySamplingBased2021,seddonQuantifyingQuantumSpeedups2021}.
The central idea of quasiprobability simulators is to decompose a complex operator (operation) $A$ over a discrete set of classically tractable operators (operations) $\Bqty{B_i}$, i.e., $A=\sum_i q_i B_i$.
The examples of the classically tractable operators $\Bqty{B_i}$ are pure stabilizer states \cite{howardApplicationResourceTheory2017,hakkakuComparativeStudySamplingBased2021} and Pauli operators \cite{rallSimulationQubitQuantum2019}.
The coefficients of the decomposition $\Bqty{q_i}$ are called ``quasiprobability distribution'', and the L1 norm of the quasiprobability distribution $\sum_i \abs{q_i}$ determines a sampling cost.
Any set of operations that can be efficiently simulated by classical computers can be used as operators $\Bqty{B_i}$ in a quasiprobability-based simulator.
Over the past years, Clifford circuits have become a popular class of such channels.~\cite{howardApplicationResourceTheory2017,hakkakuComparativeStudySamplingBased2021,seddonQuantifyingMagicMultiqubit2019,seddonQuantifyingQuantumSpeedups2021}.

Here, we consider another popular class of classically simulatable circuits: fermionic linear optics (FLO) and matchgates \cite{knillFermionicLinearOptics2001,terhalClassicalSimulationNoninteractingfermion2002,bravyiLagrangianRepresentationFermionic2005,jozsaMatchgatesClassicalSimulation2008}.
These classes represent the dynamics of free fermions, generated by quadratic fermionic Hamiltonians.
It represents a restricted class of quantum circuits in the sense that, in general, natural fermionic interactions are described by not only two but also four fermionic modes.
For example, the four fermionic modes appear in quantum circuits tailored to simulate fermionic systems, such as unitary coupled cluster (UCC) \cite{peruzzoVariationalEigenvalueSolver2014,grimsleyAdaptiveVariationalAlgorithm2019} and Hamiltonian variational {\ansatze} \cite{weckerProgressPracticalQuantum2015,wiersemaExploringEntanglementOptimization2020}.
Such four fermionic modes make the quantum dynamics or variational circuit {\ansatze} hard to be simulated on classical computers, which provides a potential advantage of using quantum computers.
However, the required amount of the four fermionic modes is limited in a certain (but common) situation where the Hartree-Fock calculation provides a good approximation.
Hence, one may be able to classically simulate quantum circuits for such systems.

This paper presents a quasiprobability-based simulator exploiting FLO and the corresponding measure that quantifies the sampling cost.
This simulator decomposes a fermionic non-Gaussian operation, such as a four fermionic modes, over all possible free operations of FLO.
The simulation cost is characterized by ``fermionic nonlinearity,'' defined as the minimum of the L1 norm of the quasiprobability distribution.
We calculate the upper bound of fermionic nonlinearity of a four fermionic modes under stochastic Pauli noise as an example, thereby estimating the sampling cost to simulate famous VQE {\ansatze} for fermionic Hamiltonian.
More specifically, we estimate the sampling cost of the noisy UCCSD quantum circuits for the hydrogen chain up to H$_8$ with several spacings using the optimized variational parameters obtained by full-vector simulations.
A rough extrapolation from the results shows us that the noisy UCCSD quantum circuits for the hydrogen chain with the spacing of \SI{0.8}{\angstrom} at the error rate of the two-qubit dephasing noise $p=0.02$ can be simulated up to H$_{22}$ within a reasonable sampling cost.
Furthermore, if $p=0.03$, the noisy UCCSD quantum circuit for the arbitrary-length hydrogen chain can be simulated because of vanishing fermionic nonlinearity.
We also study the overhead of a quantum error mitigation method when applying it to noisy UCCSD quantum circuits and compare the overheads with the sampling costs of simulating error-free UCCSD quantum circuits by our proposed method.
We believe that our method and results are helpful to design quantum circuits that simulate fermionic systems for potential quantum supremacy or quantum advantages.

\section{Fermionic Nonlinearity of Quantum Circuits}
\NewDocumentCommand\mfo{}{\hat{c}} 
\NewDocumentCommand\fao{}{\hat{a}} 
\NewDocumentCommand\fco{}{\hat{a}^\dag} 
\NewDocumentCommand\nf{}{W} 
\subsection{Definition of fermionic nonlinearity}
Here we briefly review the efficient simulatability of FLO before explaining the quasiprobability method to simulate general fermionic interactions through FLO.
We define $\Bqty{\mfo_i}_{i=1}^{2n}$ as the Majorana fermion operators that satisfy
\begin{align}
    \acomm{\mfo_i}{\mfo_j} = 2\delta_{ij}, \\
    \mfo^\dag_i = \mfo_i,                  \\
    \mfo_i^2 = I.
\end{align}
The fermionic covariance matrix of a (unnormalized) mixed state $\rho$ is defined as
\begin{align}
    M_{ij} = \frac{i}{2}\frac{\Tr(\rho\comm{\mfo_i}{\mfo_j})}{\Tr(\rho)}.
\end{align}
We call $\rho$ a fermionic Gaussian state (FGS) iff its covariance matrix $M$ satisfies $MM^T=I$.
An FGS is fully specified by the covariance matrix and the norm $\Gamma=\Tr(\rho)$.
An operator $G$ is called a fermionic Gaussian operator (FGO) iff it maps an FGS to an FGS by conjugation.
An arbitrary FGO can be written in the form of $\exp{\sum_{i<j} g_{ij} \mfo_i\mfo_j}$.
It is known that the evolution of an FGS under an FGO can be efficiently simulated on classical computers.
See Refs.~\cite{bravyiLagrangianRepresentationFermionic2005,bravyiEfficientAlgorithmsMaximum2014} for the details.

We now describe a quasiprobability-based simulation method of general fermionic interactions via FLO.
Let $\mathcal{S}_i$ be a map defined as follows:
\begin{align}
 \mathcal{S}_i \pqty{\rho} \coloneqq U_i \rho V_i^\dag,
\end{align}
where $U_i$ and $V_i$ are trace-preserving FGOs, and let $\Bqty{\mathcal{S}_i}$ be a set of all possible $\mathcal{S}_i$.
Given a fermionic non-Gaussian completely positive trace-preserving (CPTP) channel $\mathcal{E}$
, we seek to express $\mathcal{E}$ as,
\begin{align}
    \mathcal{E} &= \sum_i q_i \mathcal{S}_i \\
    &= \sum_i p_i \norm{q}_1 e^{i\theta_i} \mathcal{S}_i, \label{eq:quasiprobability}
\end{align}
where
\begin{align}
    \norm{q}_1 =& \sum_i\abs{q_i},\\
    p_i \coloneqq& \frac{\abs{q_i}}{\norm{q}_1},\\
    e^{i\theta_i} =& \frac{q_i}{\abs{q_i}}.
\end{align}
$p_i$ is a probability distribution because $p_i$ is non-negative and sum to unity.
If this decomposition can be made, we can simulate the $\mathcal{E}$ by sampling a $\mathcal{S}_i$ with probability $p_i$ and multiplying the coefficient $\norm{q}_1e^{i\theta_i}$ to the results afterwards.
The square of the L1 norm $\norm{q}_1$ quantifies the classical simulation cost of this Monte-Carlo type simulation.

To be more concrete, 
let us consider the expectation value of a two-body fermionic interaction $\mfo_\mu \mfo_\nu$ with respect to $\mathcal{E}(\rho)$ for an FGS state $\rho$ and a fermionic non-Gaussian CPTP channel $\mathcal{E}$.
The desired quantity can be written as,
\begin{align}
    \ev{\mfo_\mu \mfo_\nu} &= \Tr(\mathcal{E}\pqty{\rho}\mfo_\mu \mfo_\nu)\\
    &= \sum_i p_i \norm{q}_1\Re{e^{i\theta_i} \ev{\mfo_\mu \mfo_\nu}_{\mathcal{S}_i}} \label{eq:decomposition_over_FLO}
\end{align}
where 
\begin{align}
\ev{\mfo_\mu \mfo_\nu}_{\mathcal{S}_i} \coloneqq \Tr(\mathcal{S}_i\pqty{\rho}\mfo_\mu \mfo_\nu),
\end{align}
and $\ev{\mfo_\mu \mfo_\nu}_{\mathcal{S}_i}$ is complex in general.
From \cref{eq:decomposition_over_FLO}, we can calculate the desired quantity $\ev{\mfo_\mu \mfo_\nu}$ by sampling an index $i$ with probability $p_i$ and calculating $\norm{q}_1 \Re{e^{i\theta_i}\ev{\mfo_\mu \mfo_\nu}_{\mathcal{S}_i}}$ efficiently.
$\ev{\mfo_\mu \mfo_\nu}_{\mathcal{S}_i}$ can be calculated efficiently if $\rho$ can be written by a combination of pure FGOs, i.e., $\rho=\sum_i \lambda_i \op{\psi_\text{FGS $i$}}$.
This is because $\ev{\mfo_\mu \mfo_\nu}_{\mathcal{S}_i} = \sum_i \lambda_i \mel{\psi_\text{FGS $i$}}{V_i^\dag \mfo_\mu \mfo_\nu U_i}{\psi_\text{FGS $i$}}$, where FLO can simulate $U_i\ket{\psi_\text{FGS $i$}}$ and $V_i\ket{\psi_\text{FGS $i$}}$ classically efficiently.
Let $N$ be the number of samples.
Then, $\norm{q}_1\Re{e^{i\theta_i}\ev{\mfo_\mu \mfo_\nu}_{\mathcal{S}_i}}$ is an unbiased estimator of the desired quantity $\ev{\mfo_\mu \mfo_\nu}$.
$\norm{q}_1\Re{e^{i\theta_i}\ev{\mfo_\mu \mfo_\nu}_{\mathcal{S}_i}}$ is bounded in the interval $\bqty{-\norm{q}_1, \norm{q}_1}$, and thus the Hoeffding inequality \cite{hoeffdingProbabilityInequalitiesSums1963} shows that to estimate $\ev{\mfo_\mu \mfo_\nu}$ within additive error at most $\epsilon$ with probability at least $1-\delta$, we must set the number of samples such that
\begin{align}
    N \geq 2\norm{q}_1^2\frac{1}{\epsilon^2}\ln{\frac{2}{\delta}} \label{eq:samples_fn}.
\end{align}
Note that the expectation value of higher-order correlation function can be estimated in a similar way exploiting Wick's theorem.

Having seen that $\norm{q}_1$ determines the sampling cost for simulations, we define the fermionic nonlinearity of a quantum channel $\mathcal{E}$ as follows:
\begin{align}
\nf\pqty{\mathcal{E}} \coloneqq \min_{\Bqty{q_i}}\Bqty{\norm{q}_1|\mathcal{E} = \sum_i q_i \mathcal{S}_i}. \label{eq:f_nonlinearity}
\end{align}
$\nf\pqty{\mathcal{E}}$ quantifies the minimum number of samples to execute the Monte-Carlo type simulation of a quantum circuit with FLO.
Moreover, fermionic nonlinearity is submultiplicative under composition, i.e., $\nf\pqty{\mathcal{E}_2 \circ \mathcal{E}_1}\leq \nf\pqty{\mathcal{E}_2}\nf\pqty{\mathcal{E}_1}$.
This property helps to estimate the upper bound of the sampling cost of an $n$-mode fermionic quantum channel when $n$ is too large to calculate the fermionic nonlinearity directly.
We prove the submultiplicativity in Appendix \ref{sec:submultiplicativity}.

\subsection{Concrete decomposition for four fermionic modes}
So far, we have assumed that there exists a decomposition of non-FGO in the form of \cref{eq:quasiprobability}.
Here we consider how to explicitly calculate the decomposition and the fermionic nonlinearity of four fermionic modes in the form of $\exp(i\theta \mfo_i\mfo_j\mfo_k\mfo_l)$.
This type of operation appears in the simulation of interacting fermions and in VQE {\ansatze} such as the UCC {\ansatz} \cite{peruzzoVariationalEigenvalueSolver2014}, and its variant, Jastrow-type {\ansatz} \cite{leeGeneralizedUnitaryCoupled2019,matsuzawajastrowtypeDecompositionQuantum2020}, or Hamiltonian variational {\ansatz} (HVA) \cite{weckerProgressPracticalQuantum2015,wiersemaExploringEntanglementOptimization2020}.
Without loss of generality, we consider the decomposition of the following operator,
\begin{align}
\mathcal{E}_\text{rot} = \bqty{\exp\pqty{-i\theta \mfo_{1} \mfo_{2} \mfo_{3} \mfo_{4}}},
\end{align}
where $\bqty{A} \rho \coloneqq A\rho A^\dag$.
This is because we can always perform transformations $\mfo_i\to\mfo_1$, $\mfo_j\to\mfo_2$, $\mfo_k\to\mfo_3$, and $\mfo_l\to\mfo_4$ by FGOs for mutually distinct indices $i$, $j$, $k$, and $l$.
The four fermionic modes $\mfo_{1} \mfo_{2} \mfo_{3} \mfo_{4}$ can be mapped to a Pauli operator by the Jordan-Wigner transformation.
It maps $\mfo_k$ as follows:
\begin{align}
    \mfo_{2k-1} &= X_k \prod_{j<k} Z_j,\\
    \mfo_{2k} &= Y_k \prod_{j<k} Z_j.
\end{align}
Hence, the fermionic interaction $\mathcal{E}_\text{rot}$ can be rewritten as
\begin{align}
    \mathcal{E}_\text{rot} = \bqty{\exp\pqty{i\theta Z\otimes Z}}.
\end{align}
Theoretically, the fermionic nonlinearity should be calculated using all possible FGOs.
However, in practice, it is difficult to calculate the fermionic nonlinearity using all possible FGOs because the number of all possible FGOs is infinite.
Therefore, here we calculate the upper bound of the fermionic nonlinearity using a discrete set of the FGOs.
Below, we will omit the term ``upper bound'' if there is no risk of confusion.
Furthermore, for simplicity, we use $\mathcal{S}_i$ satisfying $U_i=V_i$.
Under this condition, the estimator is simplified to $\norm{q}_1 \ev{\mfo_{\mu}\mfo_{\nu}}_{\mathcal{S}_i}$ where $\ev{\mfo_{\mu}\mfo_{\nu}}_{\mathcal{S}_i}$ is real.

We adopt the following trace-preserving fermionic Gaussian channels as the basis channels to decompose $\mathcal{E}_\text{rot}$:
\begin{align}
     &\Bqty{\bqty{e^{\pm i\frac{\pi}{4}Z}}, \bqty{I}, \bqty{Z}}^{\otimes 2}
    \cup \Bqty{\mathcal{K}_{1,\alpha},\mathcal{K}_{2,\alpha} | \alpha=\pm 1} \\ 
     \cup& \Bqty{\bqty{e^{\pm i\frac{\pi}{4}G}} | G \in \Bqty{XX, YY, XY, YX}},  \label{eq:basis_channels}
\end{align}
where
\begin{align}
    \mathcal{K}_{1,\alpha} &\coloneqq \bqty{\frac{I+Z}{2}\otimes e^{i\alpha\frac{\pi}{4}Z}} + \bqty{\frac{I-Z}{2}\otimes e^{-i\alpha\frac{\pi}{4}Z}},\\
    \mathcal{K}_{2,\alpha} &\coloneqq \bqty{e^{i\alpha\frac{\pi}{4}Z} \otimes \frac{I+Z}{2}} + \bqty{e^{-i\alpha\frac{\pi}{4}Z} \otimes \frac{I-Z}{2}}.
\end{align}
The sets of the channels in \cref{eq:basis_channels}, $\Bqty{\bqty{e^{\pm i\frac{\pi}{4}Z}}, \bqty{I}, \bqty{Z}}^{\otimes 2}$ and $\Bqty{\mathcal{K}_{1,\alpha},\mathcal{K}_{2,\alpha}}$, are adopted from Ref.~\cite{mitaraiConstructingVirtualTwoqubit2021}, where the authors provided the way to simulate a two-qubit gate, such as $\mathcal{E}_\text{rot}$, by sampling a single-qubit operation.
Aside from them, we add $\exp(\pm i\pi/4 G)$ because they are the generators of FGOs and may be used for the decomposition.
All of the elements in \cref{eq:basis_channels} are FGOs.
Indeed, $IZ$, $XX$, $XY$, $YX$, $YY$, $ZI$ can be rewritten as
\begin{gather}
    I_{k}Z_{k+1} = -i \mfo_{2k+1} \mfo_{2k+2}\qc
    X_kX_{k+1} = -i \mfo_{2k} \mfo_{2k+1}, \\
    X_{k}Y_{k+1} = -i \mfo_{2k} \mfo_{2k+2}\qc
    Y_{k}X_{k+1} = i \mfo_{2k-1} \mfo_{2k+1}, \\
    Y_kY_{k+1} = i \mfo_{2k-1} \mfo_{2k+2}\qc
    Z_kI_{k+1} = -i \mfo_{2k-1} \mfo_{2k}.
\end{gather}
Thus the exponentials of these operators in \cref{eq:basis_channels} are FGOs.
Moreover, the projective measurements in $\mathcal{K}_{i, \alpha}$ ($i=1,2$) are FGOs \cite{bravyiLagrangianRepresentationFermionic2005}.

\begin{figure}[tb]
    \centering
    \includegraphics[width=\linewidth]{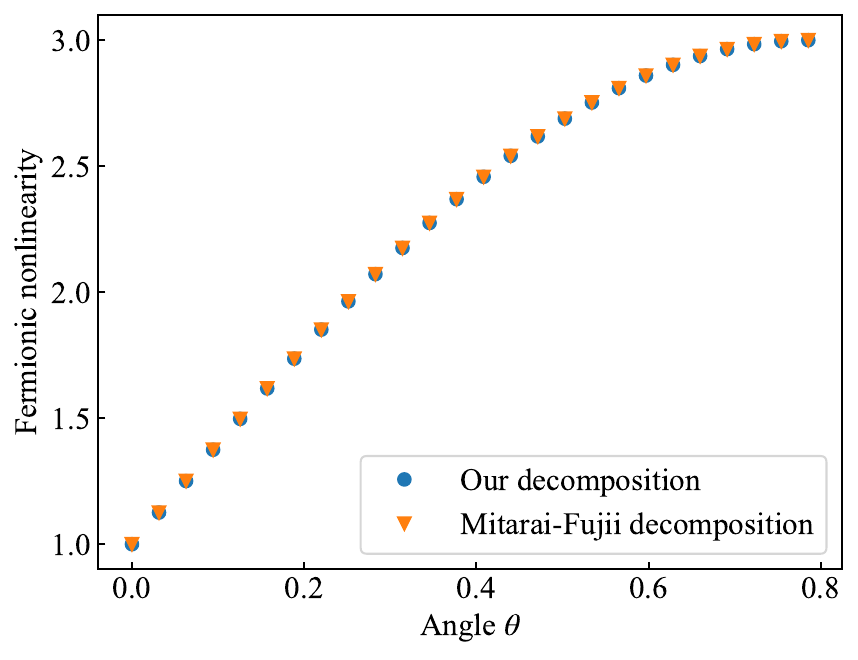}
    \caption{Fermionic nonlinearity of $\mathcal{E}_\text{rot}=\bqty{e^{i\theta \mfo_1\mfo_2\mfo_3\mfo_4}}$ as a function of the angle $\theta$.
        The horizontal axis shows the angle $\theta$.
        The vertical axis shows the fermionic nonlinearity.
        The blue circle illustrates the fermionic nonlinearity obtained by the basis channels in \cref{eq:basis_channels}.
        The orange triangle shows the fermionic nonlinearity obtained by the decomposition in Ref.~\cite{mitaraiConstructingVirtualTwoqubit2021}.
    }
    \label{fig:ch_robustnes_fgu}
\end{figure}
We calculate the fermionic nonlinearity of $\mathcal{E}_\text{rot}$ by solving the minimization problem in \cref{eq:f_nonlinearity} using the basis channels in \cref{eq:basis_channels}.
To calculate fermionic nonlinearity, we use a convex-optimization solver CVXPY \cite{diamondCVXPYPythonEmbeddedModeling2016,agrawalRewritingSystemConvex2018}.
The results are shown in \cref{fig:ch_robustnes_fgu}.
Also the fermionic nonlinearity using the decomposition in Ref.~\cite{mitaraiConstructingVirtualTwoqubit2021} is shown in \cref{fig:ch_robustnes_fgu} to compare with our results.
According to \cref{fig:ch_robustnes_fgu}, the fermionic nonlinearity is the same as when one uses the decomposition in Ref. \cite{mitaraiConstructingVirtualTwoqubit2021}.
Also, we have confirmed that the generators $\exp(\pm i\pi/4 G)$ for $G\in \{XX, XY, YX, YY\}$ do not contribute to the decomposition by examining the coefficients of the decomposition.
Moreover, we have numerically checked that the fermionic nonlinearity does not decrease even if we add the basis channels whose rotation angles are changed from $\pi/4$ to $\pi/8$, or $\pi/16$ in \cref{eq:basis_channels}.
Therefore, a measurement of a qubit in $Z$ basis, $\pm \pi/2$ rotations around the $Z$ axis, and Pauli $Z$ contribute significantly to the fermionic nonlinearity.

\begin{figure}[tb]
    \centering
    \includegraphics[width=\linewidth]{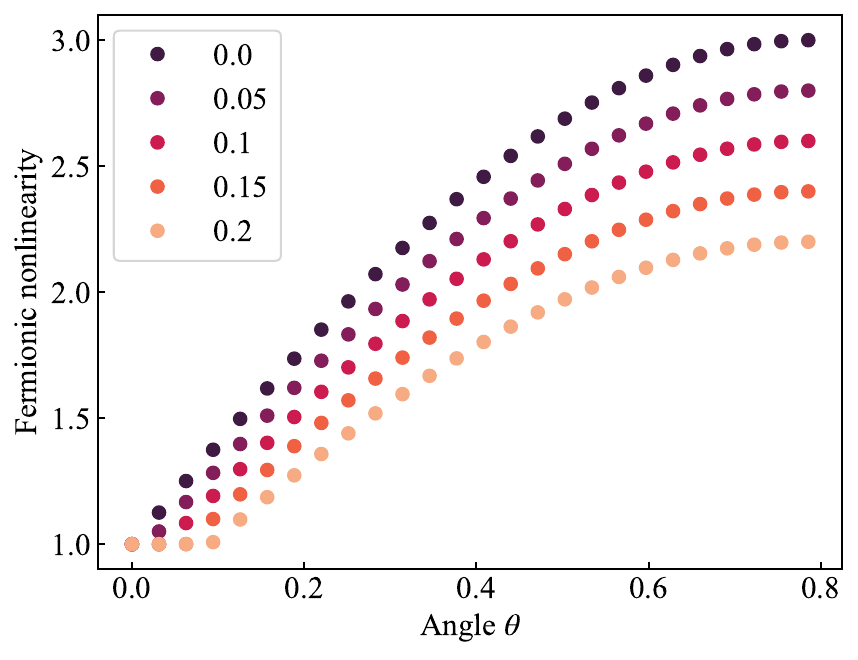}
    \caption{Fermionic nonlinearity of $\mathcal{E}_\text{noisy rot}\coloneqq \mathcal{N}_\text{dep} \circ \mathcal{E}_\text{rot}$ as a function of the angle $\theta$ of $\mathcal{E}_\text{rot}$ and the error rate $p$ of the two-qubit dephasing noise $\mathcal{N}_\text{dep}$.
        The horizontal axis shows the angle $\theta$ of $\mathcal{E}_\text{rot}$.
        The vertical axis shows the fermionic nonlinearity.
        The basis channels decomposing $\mathcal{E}_\text{noisy rot}$ are shown in \cref{eq:basis_channels}.
        The legend shows the error rate $p$ of $\mathcal{N}_\text{dep}$.
    }
    \label{fig:ch_robustness_noisy_rot_ZZ}
\end{figure}

\subsection{Fermionic nonlinearity of noisy channels and application to the VQE simulation}
We consider the fermionic nonlinearity of $\mathcal{E}_\text{rot}$ being subject to noise:
\begin{align}
    \mathcal{E}_{\text{noisy rot}} \coloneqq \mathcal{N}_{\text{dep}} \circ \mathcal{E}_{\text{rot}},
\end{align}
where $\mathcal{N}_{\text{dep}}$ is the two-qubit dephasing noise
\begin{align}
    \mathcal{N}_\text{dep} \coloneqq \pqty{1-p}\bqty{I^{\otimes 2}} + \frac{p}{3} \pqty{\bqty{IZ} + \bqty{ZI} + \bqty{ZZ}}
\end{align}
$p$ is the error rate of the dephasing noise.
\Cref{fig:ch_robustness_noisy_rot_ZZ} shows the fermionic nonlinearity of $\mathcal{E}_\text{noisy rot}$ as a function of the angle of $\mathcal{E}_\text{rot}$ and the error rate of $\mathcal{N}_\text{dep}$.
We see that the fermionic nonlinearity decreases as the error rate of $\mathcal{N}_\text{dep}$ increases.
In addition, the smaller the rotation angle, the more easily the noise makes the fermionic nonlinearity unity, that is, such a noisy fermionic interaction becomes a probabilistic mixture of FGOs.
One implication of these results is as follows.
For VQEs of fermionic problems, 
if the Hartree-Fock approach is a good first-order approximation and hence fermionic nonlinearity of the {\ansatz} stays small even after the optimization, such a quantum circuit is fragile against noise in the sense that it 
readily becomes simulatable by the proposed sampling method.

To analyze more practical cases, we estimate the sampling cost of VQE that aims to obtain the ground state of the electronic Hamiltonian of the hydrogen chain H$_m$.
Such a Hamiltonian is often used to benchmark the performance of classical quantum chemistry simulations \cite{hachmannMultireferenceCorrelationLong2006,al-saidiBondBreakingAuxiliaryfield2007,tsuchimochiStrongCorrelationsConstrainedpairing2009,sinitskiyStrongCorrelationHydrogen2010,mazziottiLargeScaleSemidefiniteProgramming2011,linDynamicalMeanFieldTheory2011,stellaStrongElectronicCorrelation2011,mottaSolutionManyElectronProblem2017,mottaGroundStatePropertiesHydrogen2020} and VQE \cite{mitaraiQuadraticCliffordExpansion2022} numerically.
This is because it exhibits rich phenomena, including metal-insulator transitions, and one can benchmark methods in both strong and weak correlation regimes.
In particular, the Hamiltonian of the hydrogen chain with the use of the STO-3G basis set has a connection with the Hubbard model; the large spacing of the hydrogen chain corresponds to the Hubbard model in the large coupling limit, and vice versa.
As for VQE, the authors of Ref.~\cite{aruteHartreeFockSuperconductingQubit2020} have demonstrated that their quantum computer can prepare the Hartree-Fock state of H$_{12}$ using VQE, although their variational {\ansatz} circuit is classically efficiently simulatable by FLO because the quantum circuit consists of two-body fermionic interactions.
In the following numerical simulation, the Hamiltonians are generated by \small{OPENFERMION} \cite{mccleanOpenFermionElectronicStructure2020} and \small{PYSCF} \cite{sunPySCFPythonbasedSimulations2018,sunRecentDevelopmentsPySCF2020} with the use of the STO-3G basis set, and then the Jordan-Wigner transformation maps them to qubit Hamiltonians, resulting in $2m$-qubit Hamiltonian for an $m$-hydrogen chain H$_m$.
We take the Hartree-Fock (HF) state $\ket{\mathrm{HF}}$ as the reference state for the VQE.

We consider the UCC {\ansatz} \cite{kutzelniggQuantumChemistryFock1982,kutzelniggQuantumChemistryFock1983,kutzelniggQuantumChemistryFock1985,bartlettAlternativeCoupledclusterAnsatze1989,kutzelniggErrorAnalysisImprovements1991,taubeNewPerspectivesUnitary2006}, which is a chemically inspired {\ansatz} and often used in VQEs \cite{peruzzoVariationalEigenvalueSolver2014,grimsleyAdaptiveVariationalAlgorithm2019}.
In particular, we consider the UCCSD {\ansatz} that only includes single and double excitations.
The UCCSD {\ansatz} is defined as
\begin{align}
    U &= e^{\pqty{T_1-T_1^\dag} + \pqty{T_2 - T_2^\dag}},\\
    T_1 &\coloneqq \sum_{a\in\text{virt}, i \in \text{occ}} t_{a i} \fco_a \fao_i, \\
    T_2 &\coloneqq \sum_{a, b \in \text{virt}, i, j \in \text{occ}} t_{ab ij} \fco_a \fco_b \fao_i \fao_j,
\end{align}
where occ and virt represent the sets of occupied and virtual orbitals, respectively, and $t_{ai}$ and $t_{abij}$ are variational parameters.
Usually, the UCCSD is implemented as a quantum circuit by Trotter expansion of $U$:
\begin{align}
\begin{split}
    \tilde{U} &= \left[\prod_{a\in\text{virt}, i \in \text{occ}} \exp\left(\frac{t_{a i}}{N_{\mathrm{trot}}} (\fco_a \fao_i-\fco_i \fao_a)\right)\right. \\
    &\quad \left. \prod_{a, b \in \text{virt}, i, j \in \text{occ}} \exp\left(\frac{t_{ab ij}}{N_{\mathrm{trot}}}( \fco_a \fco_b \fao_i \fao_j-\fco_j \fco_i \fao_b \fao_a) \right)\right]^{N_{\mathrm{Trot}}}
\end{split}
\end{align}
In the following, we consider the UCCSD {\ansatz} with $N_{\mathrm{Trot}}=1$.
Note that the fermion operator $\fao_i$ is associated with the Majorana fermion operators $\mfo_{2i-1}, \mfo_{2i}$ as follows:
\begin{align}
\mfo_{2i-1} &= \fao_i + \fco_i,\\
\mfo_{2i} &= -i\pqty{\fao_i - \fco_i}.
\end{align}
Using this relation, a four fermionic modes constituting $\tilde{U}$ can be rewritten by the Majorana fermion operators as follows:
\begin{align}
\begin{split}
&e^{t_{abij}\pqty{\fco_a \fco_b \fao_i \fao_j - \fco_j \fco_i \fao_b \fao_a}}\\
    =& e^{-i\frac{t_{abij}}{8}\mfo_{2a-1}\mfo_{2b-1}\mfo_{2i-1}\mfo_{2j}}
    e^{-i\frac{t_{abij}}{8}\mfo_{2a-1}\mfo_{2b-1}\mfo_{2i}\mfo_{2j-1}}\\
    & e^{-i\frac{t_{abij}}{8}\mfo_{2a-1}\mfo_{2b}\mfo_{2i-1}\mfo_{2j-1}}
    e^{i\frac{t_{abij}}{8}\mfo_{2a-1}\mfo_{2b}\mfo_{2i}\mfo_{2j}}\\
    & e^{-i\frac{t_{abij}}{8}\mfo_{2a}\mfo_{2b-1}\mfo_{2i-1}\mfo_{2j-1}}
    e^{i\frac{t_{abij}}{8}\mfo_{2a}\mfo_{2b-1}\mfo_{2i}\mfo_{2j}}\\
    & e^{i\frac{t_{abij}}{8}\mfo_{2a}\mfo_{2b}\mfo_{2i-1}\mfo_{2j}}
    e^{i\frac{t_{abij}}{8}\mfo_{2a}\mfo_{2b}\mfo_{2i}\mfo_{2j-1}}.\label{eq:UCCSD_double_excitation}
    \end{split}
\end{align}
We consider the sampling cost for simulating UCCSD circuits when each of the Majorana rotation gates in \cref{eq:UCCSD_double_excitation} are subjected to dephasing noise.

The sampling cost of a UCCSD quantum circuit can be given by the upper bound of the fermionic nonlinearity, which can be calculated by the product of the fermionic nonlinearity of the four fermionic modes.
Note that, as mentioned before, the HF states used as the reference states are FGSs; therefore, there are no sampling costs due to the input states.
We use the optimized variational parameters of error-free UCCSD quantum circuits, calculated by the full-vector simulations performed with Qulacs \cite{suzukiQulacsFastVersatile2021}.

\begin{figure}[tb]
    \centering
    \includegraphics[width=\linewidth]{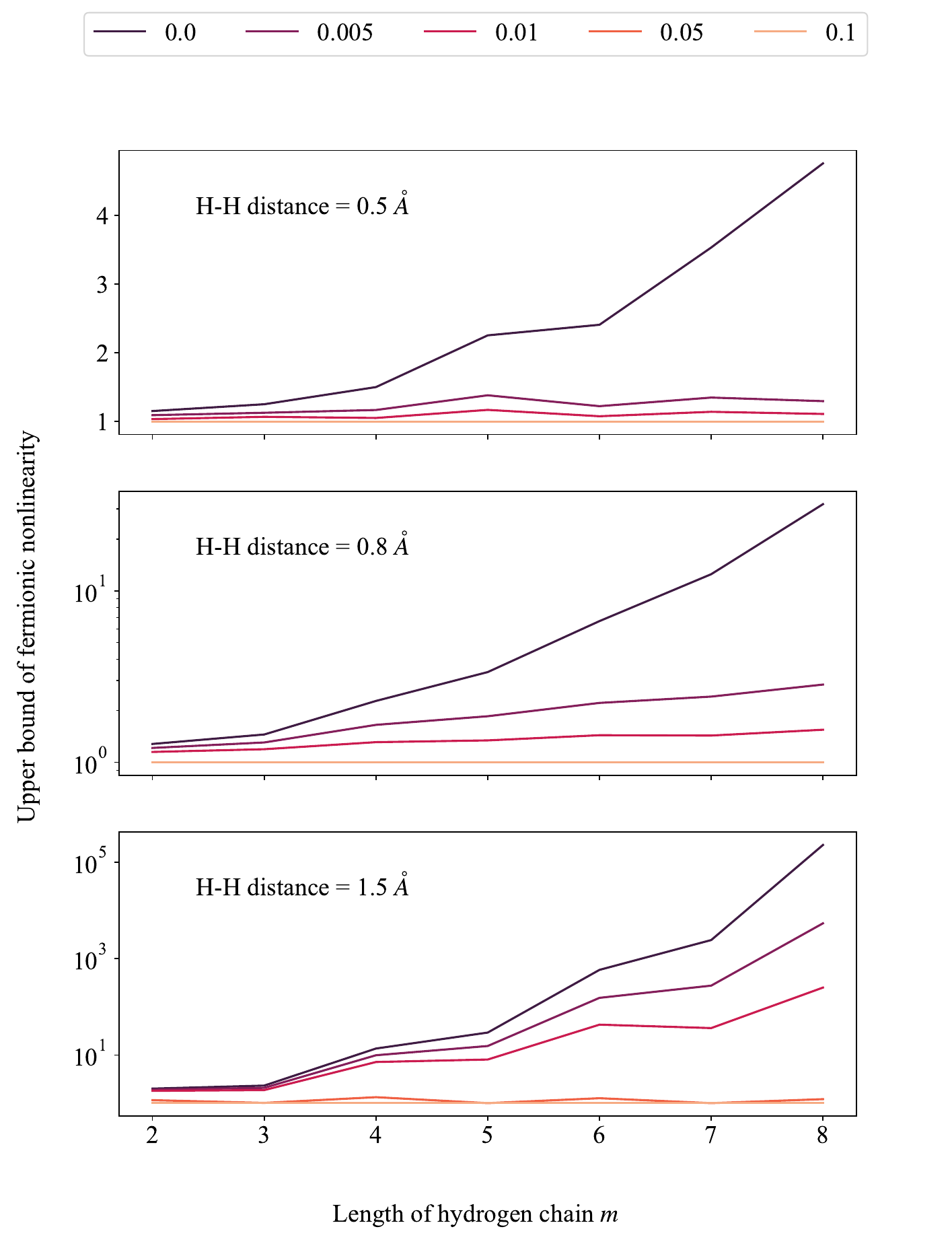}
    \caption{Upper bound of the fermionic nonlinearity of the UCCSD quantum circuit for the hydrogen chain up to H$_8$ at different spacings, \SI{0.5}{\angstrom}, \SI{0.8}{\angstrom}, and \SI{1.5}{\angstrom}.
    The horizontal axis shows the length of the hydrogen chain.
    The vertical axis shows the upper bound of the fermionic nonlinearity.
    The legend shows the error rate $p$ of the dephasing noise $\mathcal{N}_\text{dep}$.
    }
    \label{fig:cost_UCCSD}
\end{figure}
\Cref{fig:cost_UCCSD} shows the upper bound of fermionic nonlinearity of the UCCSD quantum circuit as a function of the length $m$ of the hydrogen chains H$_m$ and the error rate of the dephasing noise at different spacings of the hydrogen atoms.
In the case of \SI{0.5}{\angstrom} and \SI{0.8}{\angstrom}, the Hamiltonian for the hydrogen chain embodies a weakly correlated electronic system.
In contrast, the Hamiltonian in the case of \SI{1.5}{\angstrom} provides a strongly correlated electronic system.
From \cref{fig:cost_UCCSD}, we find that the fermionic nonlinearity is smaller when the spacing of the hydrogen chain is smaller.
This reflects that an HF state is a good approximation of the ground state when the spacing is small.
Note that the upper bounds may be overestimated since we use the submultiplicativity of fermionic nonlinearity.

Next, we discuss the size of the hydrogen chain that can be simulated within one day using $10^6$ CPU cores.
Suppose that we want to estimate the expectation value of the Hamiltonian $H$ for the hydrogen chain within an additive error $\epsilon = \norm{H}_\text{op} 10^{-3}$, with a success probability of at least $1-\delta = 1- 10^{-2}$, where $\norm{A}_\text{op}$ is the operator norm of $A$.
Assuming that each core takes at most 1 ms to calculate one sample of a quasiprobability distribution, a UCCSD quantum circuit whose fermionic nonlinearity is upper-bounded by $3 \times 10^3$ could be simulated within one day.
We estimate the fermionic nonlinearity for $m > 8$ using the geometrical mean of fermionic nonlinearity of $\mathcal{E}_\text{noisy rot}$ in the noisy UCCSD quantum circuit at $m=8$, $\overline{W\pqty{\mathcal{E}_\text{noisy rot}}}_{m=8}$. 
Let $N_4$ be the total number of four fermionic modes in $\hat{T}_2-\hat{T}_2^\dagger$.
We estimate the upper bound of fermionic nonlinearity of H$_m$ ($m>8$) by
\begin{align}
    \bqty{\overline{W\pqty{\mathcal{E}_\text{noisy rot}}}_{m=8}}^{N_4}.
\end{align}
At $p=0.02$ and the spacing of \SI{0.8}{\angstrom}, $\overline{W\pqty{\mathcal{E}_\text{noisy rot}}}_{m=8}$ is $1.00012$.
Therefore, we estimate the UCCSD circuits for hydrogen chains under such conditions can be simulated up to $m=22$ if this mean stays at the same level at larger $m$.
Furthermore, we find $\overline{W\pqty{\mathcal{E}_\text{noisy rot}}}_{m=8}=1$ if $p\geq 0.03$ and the spacing is less than \SI{0.8}{\angstrom}, or if $p\geq 0.07$ and the spacing is \SI{1.5}{\angstrom}.
We hence expect that the UCCSD circuits under such conditions  can be simulated for arbitrary $m$.

Note that the energy expectation value obtained from simulations of noisy UCCSD circuits is slightly biased from the true value $\mel{\mathrm{HF}}{U^\dag H U}{\mathrm{HF}}$.
If we allow such a bias, we can take an alternative approach; we can utilize the classical coupled cluster (CC) theory to simulate UCC circuits.
It is known that the CC theory can simulate up to large UCC systems with a small perturbative error when $t_{ab ij}$ (and $t_{a i}$) are small (i.e., small rotation angles $\theta$).
The conventional CC can be solved in polynomial time using a non-variational projection method, assuming the Hartree-Fock state to be a good reference wave function. If this assumption holds, the accuracy of the non-variational CC is almost as good as that of variational UCC ~\cite{cooperBenchmarkStudiesVariational2010,evangelistaAlternativeSinglereferenceCoupled2011}.
An established way to diagnose the correctness of the premise is by examining the magnitude of the parameters of the CC wave function ~\cite{leeDiagnosticDeterminingQuality1989, janssenNewDiagnosticsCoupledcluster1998, nielsenDoublesubstitutionbasedDiagnosticsCoupledcluster1999, leiningerNewDiagnosticOpenshell2000, leeComparisonT1D12003, fogueriSimpleDFTbasedDiagnostic2012}.
According to the rule of thumb in the classical CC, the maximum $t_{ab ij}$ is about $0.1$ or less in the region where non-variational CC works well  ~\cite{fogueriSimpleDFTbasedDiagnostic2012}.
Besides, in systems where classical CC fails, the maximum $t_{ab ij}$ tends to be larger than $0.15$ ~\cite{fogueriSimpleDFTbasedDiagnostic2012}.
Our UCC calculations show that for $m=8$, the maximum $t_{ab ij}$ is about $0.08$ when the distance between the hydrogens is \SI{0.8}{\angstrom} and about $0.18$ when the distance is \SI{1.5}{\angstrom}.
Therefore, our results are in line with the empirical trend in classical computing.

\subsection{Comparison with the overhead of probabilistic error cancellation}
Not only the classical simulatability of quantum circuits, but also whether the outputs of noisy quantum computers are accurate is important for practical applications.
Quantum error mitigation techniques are designed to reduce the bias of the outputs from noisy quantum devices \cite{liEfficientVariationalQuantum2017,temmeErrorMitigationShortDepth2017,endoPracticalQuantumError2018,mccleanHybridQuantumclassicalHierarchy2017,mcardleErrorMitigatedDigitalQuantum2019,bonet-monroigLowcostErrorMitigation2018,yoshiokaGeneralizedQuantumSubspace2022}.
One of the notable examples of quantum error mitigation is the probabilistic error cancellation (PEC) \cite{temmeErrorMitigationShortDepth2017,endoPracticalQuantumError2018}, which is a quasiprobability method.
It mitigates the effect of noise using quasiprobability at the cost of required samples to ensure a specific accuracy.
Here we compare the sampling cost of the PEC applied to noisy UCCSD quantum circuits with classical sampling costs of error-free UCCSD quantum circuits.
The central idea of PEC is that an ideal unitary $\mathcal{U}$ is represented by a linear combination of noisy implementable operations $\mathcal{V}_i$, i.e., $\mathcal{U} = \sum_i c_i \mathcal{V}_i$, where $c_i$ satisfies $\sum_i c_i = 1$, and $c_i$ is real but can be negative.
Then, we can estimate an expectation value of an observable $\ev{A}$ by sampling $\mathcal{V}_i$ with probability $\abs{c_i}/\sum_i\abs{c_i}$ and calculating $\gamma \text{sign} \pqty{c_i}\Tr\pqty{\mathcal{V}_i\pqty{\rho}A}$ many times, where $\gamma \coloneqq \norm{c}_1 = \sum_i \abs{c_i}$.
The Hoeffding inequality shows that to estimate $\ev{A}$ within additive error at most $\epsilon$ with probability at least $1-\delta$, we must set the required samples $N_\text{PEC}$ such that
\begin{align}
    N_\text{PEC}\geq 2\gamma^2 \frac{1}{\epsilon^2} \ln\frac{2}{\delta} \label{eq:samples_PEC}.
\end{align}
Thus $\gamma$ characterizes the overhead of PEC.

Let us consider the total sampling cost of PEC for a noisy UCCSD circuit being subject to the two-qubit dephasing noise, $N_\text{tot PEC}$.
For simplicity, we assume that the total number of the two-qubit dephasing noise equals the total number of four fermionic modes rotations $N_4$.
Under this assumption, $N_\text{tot PEC}$ can be given by
\begin{align}
    N_\text{tot PEC} = 2\bqty{\gamma_\text{dep}}^{2N_4} \frac{1}{\epsilon^2} \ln\frac{2}{\delta} \label{eq:number_tot_PEC},
\end{align}
where $\gamma_\text{dep}$ is the L1 norm of the coefficients of the decomposition of the PEC for two-qubit dephasing noise.
On the other hand, the classical sampling cost of the error-free UCCSD quantum circuits for H$_8$, $N_\text{tot classical}$, is given by
\begin{align}
    N_\text{tot classical} = 2\bqty{\overline{W(\mathcal{E}_\text{rot})}_{m=8}}^{2N_4} \frac{1}{\epsilon^2} \ln\frac{2}{\delta} \label{eq:number_tot_classical}.
\end{align}
From \cref{eq:number_tot_PEC,eq:number_tot_classical}, we obtain
\begin{align}
    \frac{N_\text{tot PEC}}{N_\text{tot classical}} = \bqty{\frac{\gamma_\text{dep}}{\overline{W\pqty{\mathcal{E}_\text{rot}}}_{m=8}}}^{2N_4}.
\end{align}
Thus, we can compare the sampling costs of the PEC with that of the error-free UCCSD quantum circuits by calculating $\gamma_\text{dep}$ and $\overline{W\pqty{\mathcal{E}_\text{rot}}}_{m=8}$.
$\gamma_\text{dep}$ can be given as
\begin{align}
    \gamma_\text{dep}=\frac{3+2p}{3-4p},
\end{align}
which we explain in detail in Appendix \ref{sec:overhead_PEC_dep}.

\begin{figure}[tb]
    \centering
    \includegraphics[width=\linewidth]{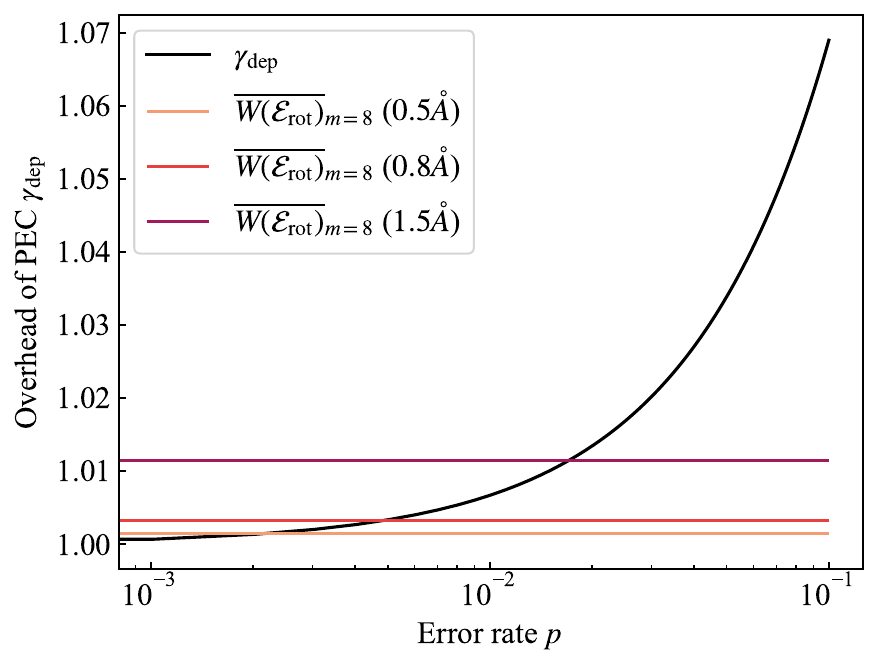}
    \caption{Overhead of the PEC of the two-qubit dephasing noise as a function of the error rate.
    The horizontal axis shows the error rate of the two-qubit dephasing noise.
    The vertical axis shows the overhead of the PEC for the two-qubit dephasing noise, $\gamma_\text{dep}$.
    To compare the overhead with classical sampling costs of error-free UCCSD quantum circuits, we also plot $\overline{W\pqty{\mathcal{E}_\text{rot}}}_{m=8}$ of UCCSD quantum circuits for H$_8$ at different spacings, \SI{0.5}{\angstrom}, \SI{0.8}{\angstrom}, and \SI{1.5}{\angstrom} as violet, red, and orange horizontal lines, respectively.
    }
    \label{fig:cost_PEC}
\end{figure}
\Cref{fig:cost_PEC} shows the overhead of the PEC for the two-qubit dephasing noise, $\gamma_\text{dep}$ as a function of the error rate of the noise, $p$.
We also plot the geometrical means of the upper bounds of fermionic nonlinearity $\overline{W\pqty{\mathcal{E}_\text{rot}}}$ of UCCSD quantum circuits for H$_8$ at the different spacings, \SI{0.5}\AA, \SI{0.8}\AA, and \SI{1.5}\AA, as horizontal lines.
We observe that there are crossovers between the overhead of the PEC and the upper bounds of the fermionic nonlinearity.
At the spacing of \SI{0.5}\AA, the crossover appears around $p=0.0021$.
In the case of the spacing of \SI{1.5}\AA, the crossover appears around $p=0.017$.
Above these threshold error rates, the PEC overheads are greater than the upper bounds of the fermionic nonlinearity, and thus noisy quantum devices with the PEC would be useless.
Otherwise, i.e., in the small error regime, the PEC overheads are less than the upper bounds of the fermionic nonlinearity, and thus noisy quantum devices with PEC would be preferred to simulate noisy UCCSD quantum circuits.

Our results indicate that, to demonstrate the quantum supremacy or quantum advantages with the UCCSD {\ansatz}, one has to choose target Hamiltonians that exhibit strong electronic correlations and execute the quantum circuits with sufficiently low error rates.
Note that even if the gate error of a device is 1\%, the effective physical error rate of the two-qubit dephasing noise in $\mathcal{E}_\text{noisy rot}$ would be much higher because, in general, the noise on the entangling gates 
to simulate non-local two or four fermionic modes rotations by
physically allowed operations accumulates.

We note that we have discussed the classical simulatability of noisy UCCSD {\ansatz} using the optimized variational parameters obtained by the error-free simulations and found that, under certain circumstances, they become classically simulatable.
In such cases, even if sophisticated error mitigation and optimization strategies allowed us to perform the VQE successfully, we cannot achieve a quantum advantage because the resulting circuit can be simulated classically.

\section{Conclusion}
In this work, we propose a quasiprobability-based simulation algorithm using FLO and quantify its simulation cost by establishing the corresponding measure, fermionic nonlinearity $W(\mathcal{E})$.
The sampling cost of the quasiprobability-based simulator is proportional to $W\pqty{\mathcal{E}}^2$.
As an example, we calculate the upper bound of fermionic nonlinearity of the noisy rotation gate generated by four fermionic modes, which often appear in the parametrized quantum circuits in VQE.
We find that the fermionic nonlinearity increases as the rotation angle becomes larger and decreases as the error rate of the dephasing noise increases.
Based on the above observation, we discuss the simulatability of the quantum circuits for quantum chemistry with our proposed method.
We estimate the sampling costs of the noisy UCCSD quantum circuits for the hydrogen chain, and discuss whether they can be simulated within one day when $10^6$ CPU cores are available.
We find that the UCCSD circuits with the dephasing error rate $p=0.02$ for hydrogen chains with the spacing of \SI{0.8}{\angstrom} can be simulated up to H$_{22}$.
Furthermore, if $p\geq0.03$, the noisy UCCSD circuits for hydrogen chain of arbitrary length with the same spacing can be simulated.
Aside from the classical simulatability of quantum circuits, the accuracy of quantum computation is also crucial.
We compare the overhead of the PEC for the two-qubit dephasing noise with that of simulating error-free UCCSD quantum circuits for the hydrogen chain classically.
We find that the noisy UCCSD quantum circuits for hydrogen chain at \SI{0.5}{\angstrom} cost more overhead than classical computers if $p\geq 0.0021$.
This analysis reveals the quantum advantage regime more clearly.
Although this numerical result is pessimistic, it stimulates to investigate or design another VQE {\ansatz} that retains quantumness against noise with the use of our results and method.

Our work leaves several open questions.
Although we use the basis channels based on  Ref.~\cite{mitaraiConstructingVirtualTwoqubit2021} to decompose the four fermionic modes, there may exist more optimal basis channels.
It is an interesting and nontrivial problem to choose the optimal discrete set of FGOs to decompose a given non-FGO.
We also expect that one could give lower bounds for fermionic nonlinearity using the technique to derive the lower bounds of PEC in Ref.~\cite{takagiOptimalResourceCost2021} and find the exact fermionic nonlinearity of a specific quantum channel.
Although we only consider UCCSD quantum circuits as a practical case in our paper, it would be interesting to analyze the classical simulatability of another VQE {\ansatz} (e.g., Hamiltonian variational {\ansatz}) or a dynamics of a fermionic Hamiltonian by our proposed method.
We formulate the classical simulatability of a quantum circuit for fermionic Hamiltonians in the channel picture, but it is also of great interest to establish such a formulation in the state picture, which should be compatible with the results shown in Ref.~\cite{hebenstreitAllPureFermionic2019}.

\begin{acknowledgments}
We are grateful for useful discussions with Yasunari Suzuki, Suguru Endo, and Ryuji Takagi.
KM is supported by JST PRESTO Grant No.~JPMJPR2019 and JSPS KAKENHI Grant No.~20K22330.
WM is supported by JST PRESTO Grant No.~JPMJPR191A and JSPS KAKENHI Grant No.~18K14181.
KF is supported by JST ERATO Grant No.~JPMJER1601, JSPS KAKENHI Grant No.~16H02211,
and JST CREST Grant No.~JPMJCR1673.
This work is supported by MEXT Quantum Leap Flagship Program (MEXT QLEAP) Grant No.~JPMXS0118067394, JPMXS0120319794, and JST Moonshot R\&D Grant No.~JPMJMS2061.
We also acknowledge support from JST COI-NEXT program.
\end{acknowledgments}

\appendix
\section{Submultiplicativity of Fermionoic Nonlinearity}\label{sec:submultiplicativity}
Let us prove the submultiplicativity of fermionic nonlinearity: $W(\mathcal{E}_1 \circ \mathcal{E}_2) \leq W(\mathcal{E}_1) W(\mathcal{E}_2)$.
Let $\mathcal{E}_1$ and $\mathcal{E}_2$ have decompositions over elements of $\Bqty{\mathcal{S}_i}$
\begin{align}
    \mathcal{E}_1 &= \sum_i x_i \mathcal{S}_i, \\
    \mathcal{E}_2 &= \sum_j y_j \mathcal{S}_j. \\
\end{align}
We consider the composition of both channels
\begin{align}
    \mathcal{E}_1 \circ \mathcal{E}_2 = \sum_{i,j} x_i y_j \mathcal{S}_i \circ \mathcal{S}_j.
\end{align}
Since the composition of FGOs is also an FGO, this gives a decomposition of $\mathcal{E}_1 \circ \mathcal{E}_2$ over elements of $\Bqty{\mathcal{S}_i}$.
Taking the absolute sum, we have
\begin{align}
    \sum_{i,j}\abs{x_iy_j} &\leq \pqty{\sum_i\abs{x_i}}\pqty{\sum_j\abs{y_j}} \\
    &= W\pqty{\mathcal{E}_1} W\pqty{\mathcal{E}_2}.
\end{align}
Therefore, $W(\mathcal{E}_1 \circ \mathcal{E}_2) \leq W(\mathcal{E}_1) W(\mathcal{E}_2)$.

\section{Overhead of probabilistic error cancellation for two-qubit dephasing noise}
\label{sec:overhead_PEC_dep}
Here we give a decomposition of an ideal two-qubit unitary $\mathcal{U}$ for the two-qubit dephasing noise $\mathcal{N}_\text{dep}$, thereby showing the overhead of the probabilistic error cancellation for the two-qubit dephasing noise.
With reference to Ref.~\cite{takagiOptimalResourceCost2021}, we consider a specific decomposition of the identity map $\mathcal{I}$ involving $\mathcal{N}_\text{dep}$:
\begin{align}
    \mathcal{I} = \frac{3-p}{3-4p} \mathcal{N}_\text{dep} \circ \mathcal{I} - \frac{p}{3-4p} \mathcal{N}_\text{dep} \circ \pqty{\bqty{IZ}+\bqty{ZI}+\bqty{ZZ}}.
\end{align}
By applying $\mathcal{U}$ to both sides from the right, we obtain
\begin{align}
    \mathcal{U} = \frac{3-p}{3-4p} \mathcal{N}_\text{dep} \circ \mathcal{U} - \frac{p}{3-4p} \mathcal{N}_\text{dep} \circ \pqty{\bqty{IZ}+\bqty{ZI}+\bqty{ZZ}} \circ \mathcal{U}\label{eq:decomposition_of_dep}.
\end{align}
From \cref{eq:decomposition_of_dep}, for any two-qubit unitary gate $\mathcal{U}$ and $0\leq p \leq \frac{3}{4}$, the overhead of PEC for $\mathcal{N}_\text{dep}$ is characterized by
\begin{align}
    \gamma_\text{dep} = \frac{3+2p}{3-4p}.
\end{align}
Note that there may exist a more improved decomposition than that shown in \cref{eq:decomposition_of_dep}.
\bibliography{main}
\end{document}